\documentstyle[prl,aps,twocolumn,mypsfig]{revtex}

\preprint{Submitted to {\it Physical Review Letters}}
\draft

\author{M. A. Nielsen$^{1}$\thanks{Electronic address :
	mnielsen@tangelo.phys.unm.edu} 
	{\em and}\/ 
	Isaac L. Chuang$^{2}$\thanks{Electronic address :
	ike@lanl.gov}}

\title{Programmable quantum gate arrays}

\address{\vspace*{1.2ex}
        \hspace*{0.5ex}{$^1$Center for Advanced Studies, Department
        of Physics and Astronomy, \\
        University of New Mexico, NM 87131-1156}}
\address{\vspace*{1.2ex}
        \hspace*{0.5ex}{$^2$Theoretical Astrophysics T-6 \\
	Los Alamos National Laboratory, Los Alamos, NM 87545 }}

\begin{document}

\date{\today}
\maketitle

\begin{abstract}
We show how to construct quantum gate arrays that can be programmed to
perform different unitary operations on a {\em data register},
depending on the input to some {\em program register}. It is shown that a {\em
universal quantum gate array} -- a gate array which can be programmed to
perform {\em any} unitary operation -- exists only if one allows the gate array
to operate in a probabilistic fashion. The universal quantum gate array we
construct requires an exponentially smaller number of gates than a classical
universal gate array.
\end{abstract}

\pacs{PACS No. 03.65.Bz}

\narrowtext

Quantum computers \cite{DiVincenzo95,Bennett95b,Ekert96} can perform arbitrary
unitary operations on a set of two-level systems known as {\em qubits}. These
unitary operations are usually decomposed as {\em quantum gate arrays} which
implement the desired unitary operation using a finite amount of resources.
Depending on what unitary operation is desired, different gate arrays are
used\cite{Barenco95}.

By contrast, a classical computer can be implemented as a fixed classical gate
array, into which is input a {\em program}, and {\em data}. The program
specifies the operation to be performed on the data.  A universal gate array
can be programmed to perform any possible function on the input data.

This paper addresses the question of whether it is possible to build analogous
{\em programmable} quantum gate arrays -- fixed circuits, which take as input a
quantum state specifying a {\em quantum program}, and a {\em data register},
to which the unitary operator corresponding to the quantum program is applied.

\begin{figure}
\begin{center}
{\mbox{\psfig{file=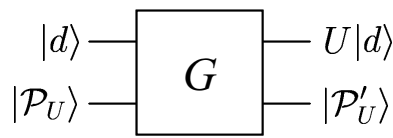,width=2.0in}}}
\end{center}
\caption{Conceptual schematic of a programmable quantum gate array which 
	implements the unitary operation $U$, determined by the quantum
	program $|{\cal P}_U\rangle$.}
\label{fig:basicidea}
\end{figure}

These gate arrays are modeled in the following manner: the initial state of
the system is assumed to be of the form
\begin{equation}
	|d\rangle \otimes |{\cal P}\rangle \,, 
\end{equation}
where $|d\rangle$ is a state of the $m$-qubit data register, and $|{\cal
P}\rangle$ is a state of the $n$-qubit program register.  Note that the two
registers are not entangled.  The total dynamics of the programmable gate array
is given by a unitary operator, $G$,
\begin{equation}
	|d\rangle \otimes |{\cal P}\rangle 
		\rightarrow G \left[ |d\rangle \otimes |{\cal P}\rangle \right]
\,.
\end{equation}
This operation is implemented by some fixed quantum gate array.  A unitary
operator, $U$, acting on $m$ qubits, is said to be {\em implemented} by this
gate array if there exists a state $|{\cal P}_U\rangle$ of the program
register such that
\begin{equation}
	G \left[ |d\rangle \otimes |{\cal P}_U\rangle \right] 
	= (U|d\rangle) \otimes |{\cal P}_U' \rangle \,, 
\end{equation}
for all states $|d\rangle$ of the data register, and some state $|{\cal
P}'_U\rangle$ of the program register. {\em A priori}, it is possible that
$|{\cal P}'_U\rangle$ depends on $|d\rangle$.  To see that this is not the
case, suppose
\begin{eqnarray}
	G \left[ |d_1\rangle \otimes |{\cal P}\rangle \right]
	= (U|d_1\rangle) \otimes |{\cal P}_1'\rangle,  \\
	G \left[ |d_2\rangle \otimes |{\cal P}\rangle \right]
	= (U|d_2\rangle) \otimes |{\cal P}_2'\rangle \,.
\end{eqnarray}
Taking the inner product of these equations we see that $\langle {\cal
P}_1'|{\cal P}_2'\rangle = 1$, and thus $|{\cal P}_1'\rangle = |{\cal
P}_2'\rangle$, and therefore there is no $|d\rangle$ dependence of $|{\cal
P}_U'\rangle$.  A schematic of this setup is shown in
Fig.~\ref{fig:basicidea}.

The set of unitary operators on $m$ qubits can be parametrized by $2^{2m}$
independent real numbers, which is fewer than the $2^{2m+1}-1$ real numbers
needed to parametrize a set of $2m$ qubits. Therefore, it seems that it
might be possible to implement a {\em universal} quantum gate array - one
which can be programmed to implement {\em any} unitary operation. Universal
gate arrays are certainly possible for classical computers, since by counting
the number of possible functions we see that an arbitrary function on $m$ bits
can be specified using $m2^m$ bits, and it is straightforward to design a
classical circuit which will take as input $m 2^m$ program bits and implement
the corresponding function on $m$ data bits.

The following result shows that no universal quantum gate array (of finite
extent) can be realized. More specifically, we show that every implementable
unitary operation requires an extra Hilbert space dimension in the program
register. Since the number of possible unitary operations on $m$ qubits is
infinite, it follows that a universal gate array would require an infinite
number of qubits in the program register, and thus no such array exists.
Note also that a program register with $d$ dimensions
can be used to implement $d$ unitary operations which are distinct up
to a global phase by performing an appropriate sequence of 
controlled unitary operations \cite{Barenco95}.

{\em Result}: Suppose distinct (up to a global phase) unitary operators
$U_1,\ldots,U_N$ are implemented by some programmable quantum gate array. Then
the program register is at least $N$-dimensional, that is, contains at least
$\log_2 N$ qubits. Moreover, the corresponding programs $|{\cal
P}_1\rangle,\ldots,|{\cal P}_N\rangle$ are mutually orthogonal.

The proof is to suppose that $|{\cal P}\rangle$ and $|{\cal Q}\rangle$ are
programs which implement unitary operators $U_p$ and $U_q$ which are distinct
up to global phase changes. Then for arbitrary data $|d\rangle$ we have
\begin{eqnarray}
	G ( |d\rangle \otimes |{\cal P}\rangle ) & = & (U_p |d\rangle )
	 \otimes |{\cal P}'\rangle \\
	G ( |d\rangle \otimes |{\cal Q}\rangle ) & = & (U_q |d\rangle )
	 \otimes |{\cal Q}'\rangle \,, 
\end{eqnarray}
where $|{\cal P}'\rangle$ and $|{\cal Q}'\rangle$ are states of the program
register. Taking the inner product of the previous two equations gives
\begin{equation}
	\langle {\cal Q}|{\cal P} \rangle =
	\langle {\cal Q}'|{\cal P}' \rangle \langle d| U_q^{\dagger}
		 U_p |d\rangle
\,. 
\label{eqtn: inner product}
\end{equation}
Suppose $\langle {\cal Q}'|{\cal P}'\rangle \neq 0$. Then dividing through both
sides of the equation gives
\begin{equation}
	\frac{\langle {\cal Q}|{\cal P} \rangle}
	{\langle {\cal Q}'|{\cal P}' \rangle} =
		\langle d| U_q^{\dagger} U_p |d\rangle \,. 
\end{equation}
The left hand side of this equation has no $|d\rangle$ dependence, and thus
$U_q^{\dagger}U_p = \gamma I$ for some c-number $\gamma$. It follows that the
only way we can have $\langle {\cal Q}'|{\cal P}'\rangle \neq 0$ is if $U_p$
and $U_q$ are the same up to a global phase. But we have assumed that
this is not so and thus $\langle {\cal Q}'|{\cal P}'\rangle = 0$.
Eq.(\ref{eqtn: inner product}) now tells us that
\begin{equation}
	\langle {\cal Q}|{\cal P}\rangle = 0 \,. 
\end{equation}
That is, the programs are orthogonal. The result follows.

This result demonstrates that no {\em deterministic}
universal quantum gate array exists. We will now see that it is possible
to implement a universal quantum gate array in a {\em probabilistic}
fashion.

\begin{figure}
\centerline{\mbox{\psfig{file=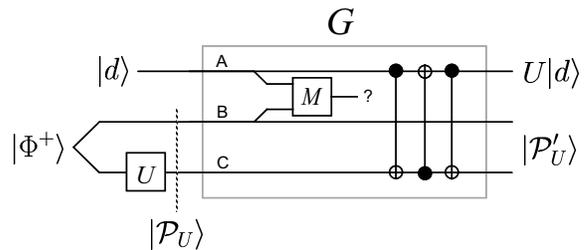,width=3.0in}}}
\caption{A probabilistic universal quantum gate array.}
\label{fig:probabilisticquga}
\end{figure}

The procedure is illustrated in Fig.~\ref{fig:probabilisticquga} for the case
of $m = 1$. In the general case the $2m$ qubit program for the $m$ qubit
unitary operation $U$ is found as follows:
\begin{equation}
	|{\cal P}_U\rangle =
	(I_m \otimes U) \, \bigotimes_{i=1}^m |\Phi^+_{i,m+i}\rangle \,, 
\end{equation}
where $I_m$ is the identity operator on the first $m$ qubits of
the program register, and the state $|\Phi^+_{x,y}\rangle$ is a Bell state
$|\Phi^+\rangle \equiv (|00\rangle + |11\rangle)/\sqrt 2$ shared between
qubits $x$ and $y$ of the program register. Joint measurements are
made on the data qubits and the first $m$ program qubits
as follows. The Bell basis is defined to consist of the states
\begin{eqnarray}
	|\Phi^{\pm}\rangle & \equiv & \frac{1}{\sqrt 2} (|00\rangle
	\pm |11\rangle )
\\	|\Psi^{\pm}\rangle & \equiv & \frac{1}{\sqrt 2} (|01\rangle
	\pm |10\rangle )
\,. 
\end{eqnarray}
Suppose a joint measurement $M$ in the Bell basis is made on the first
data qubit and the first program qubit. A joint measurement in the
Bell basis is then made on the second data qubit and the second
program qubit, and so on for all $m$ data qubits.

Specifically, for $m=1$, we have the program
\begin{equation}
	|{\cal P}\rangle 
	= (I\otimes U) |\Phi^+\rangle
	= \frac{|0\rangle U|0\rangle + |1\rangle U|1\rangle}{\sqrt 2} 
\,.
\end{equation}
For an input data register $|d\rangle = a |0\rangle + b |1\rangle$, the
input $|d\rangle |{\cal P}\rangle$ to the
gate array may be rewritten as
\begin{eqnarray}
	& & [ a |0\rangle + b |1\rangle ] \frac{ |0\rangle U|0\rangle + |1\rangle U|1\rangle }{\sqrt 2} \nonumber
\\
	&=& 	\frac{1}{2} \left[  a (|\Phi^+\rangle + |\Phi^-\rangle) U |0\rangle + \right.
%\nonumber \\
%	& & 	
		a (|\Psi^+\rangle + |\Psi^-\rangle) U |1\rangle +
\nonumber \\
	& & 	b (|\Psi^+\rangle - |\Psi^-\rangle) U |0\rangle +
%\nonumber \\
%	& &
	 	\left. b (|\Phi^+\rangle - |\Phi^-\rangle) U |1\rangle \right]
\\
	&=& 	\frac{1}{2} \left[  |\Phi^+\rangle (a U|0\rangle + b U|1\rangle) + \right.
%\nonumber \\
%	& &
	 	|\Phi^-\rangle (a U|0\rangle - b U|1\rangle) +
\nonumber \\
	& &	|\Psi^+\rangle (a U|1\rangle + b U|0\rangle) +
%\nonumber \\
%	& &
	\left. |\Psi^-\rangle (a U|1\rangle - b U|0\rangle) \right]
\\
	&=& 	\frac{1}{2} \left[  |\Phi^+\rangle [ U |d\rangle ] +
		|\Phi^-\rangle [ U \sigma_z |d\rangle ] + \right.
\nonumber \\
	& & 	\left. |\Psi^+\rangle [ U \sigma_x |d\rangle ]
		+ i |\Psi^-\rangle  [ U \sigma_y |d\rangle ] \right]
\,.
\end{eqnarray}
Now, when the measurement result from $M$ gives an eigenvalue corresponding to
$|\Phi^+\rangle$, then the post-measurement state of the second qubit of the
program register will be $U|d\rangle$, which is the desired transform. Three
controlled-NOT gates then swap the state $U|d\rangle$ of the second qubit of
the program register back into the data register, completing a successful
operation of the programmable gate array. However,
for the other three possible outcomes, the result will be different.  Thus, in
the $m=1$ case, the gate array is {\em non-deterministic}, and succeeds with
probability $1/4$.  Note that the result of the measurement tells us with
certainty whether the gate array has succeeded.

This reasoning is easily generalized to larger $m$, in which case if the result
of all the measurements corresponds to the Bell state $|\Phi^+\rangle$, then
the state of the final $m$ qubits of the program register is
$U|d\rangle$. This event has probability $2^{-2m}$, independent of the initial
state $|d\rangle$ or $U$.  To complete the operation of the universal gate
array the state of the final $m$ qubits of the program register is swapped
back into the data register, to give the desired output $U|d\rangle$.  This is
easily accomplished using cascaded controlled-NOT gates\cite{Barenco95a}.
Alternatively, the location of the data register output can be re-defined
appropriately.

Readers familiar with quantum teleportation \cite{Bennett93} can understand
why the scheme works in the following way. Divide the total system up into
three systems: $A$, the data register, $B$, the first $m$ lines of the program
register, and $C$, the final $m$ lines of the program register.  The scheme as
described is equivalent to applying $U$ to system $C$, where $B$ and $C$ are
initially bit-pairwise maximally entangled. The usual measurement procedure
for teleportation is then applied to systems $A$ and $B$. Since this procedure
involves only systems $A$ and $B$ it commutes with the application of $U$ to
system $C$, and we can suppose for the purposes of analysis that the
measurement was actually performed {\em before} the unitary $U$.  By our
knowledge of teleportation we know that for one (and only one) of the
measurement outcomes that may occur, the effect is simply to transfer the
state of system $A$ to system $C$, without the need to unitarily ``fix up''
the state of system $C$. Provided this measurement outcome, which has
probability $2^{-2m}$, occurs, the total operation is equivalent to
teleporting the data register to system $C$ and then applying $U$ to that
system.  The procedure is completed by swapping system $C$ back to system $A$.
As has been pointed out previously, this entire procedure can be accomplished
by a quantum circuit\cite{Cleve96a}.

It is clear from this explanation in terms of teleportation that the universal
gate array works for non-unitary as well as unitary quantum operations
\cite{Kraus83a,Schumacher96a}.  Unitary quantum operations have programs which
are pure states, while non-unitary operations have programs which are mixed
states.

This universal quantum gate array is particularly remarkable because the
number of gate operations is polynomial (indeed, linear) in the number of data
qubits. This is a great contrast to classical universal gate arrays, which
must be exponential in the number of data bits. To see this, consider that
there are at least $m2^m$ program bits in the classical universal gate array,
and each one of these bits must pass through at least one gate if it is to
have any effect on the data as a ``program'' bit.  If the maximum number of
bits used as input to any gate in the array is $k$, then it follows that a
classical universal gate array must have at least $m2^m/k$ gates.  The quantum
universal gate array we have demonstrated trades off an exponentially smaller
number of gates than the classical universal gate array at the expense of an
exponentially small probability of success. On {\em average} the number of
gate operations required for {\em successful operation} of the universal
quantum gate array goes like $m2^{2m}$. Where the universal quantum gate array
wins out over the classical universal gate array is the much larger variety of
transformations it is able to effect.

We have demonstrated that no deterministic universal quantum gate array
exists. More generally, a deterministic programmable gate array must have as
many Hilbert space dimensions in the program register as programs are
implemented. In the context of laboratory experiments on quantum computation,
this means that a large number of classically distinguishable states must be
available in order to build a device that can function as a general purpose
quantum computer. Fortunately, there is no shortage of such states in the
laboratory. A probabilistic universal quantum gate array has been demonstrated
that requires only a linear number of gates, but which has an exponentially
small probability of success. It would be extremely interesting to know if
this is the best that can be done, or if it is possible to build a universal
quantum gate array which is more efficient. It may also be possible to 
develop a theory of program complexity based on the universal gate
array we have proposed, perhaps based on measures of entanglement for
quantum programs.

\acknowledgements We thank Carlton M. Caves and Richard Cleve
for useful discussions about this
work.  This work was supported by the Office of Naval Research (Grant
No. N00014-93-1-0116) and the Australian-American Educational Foundation
(Fulbright Commission).

%%%%%%%%%%%%%%%%%%%%%%%%%%%%%%%%%%%%%%%%%%%%%%%%%%%%%%%%%%%%%%%%%%%%%%%%%%%%%
% References

\end{document}